\documentstyle[psfig,12pt]{article}
\begin{document}
\baselineskip=15pt
\title{ Superdeformed Bands of Odd
Nuclei in $A$=190 Region in the Quasiparticle Picture
 }
\author{ 
 J.~Terasaki, H.~Flocard \\
 Division de Physique Th\'eorique\thanks{Unit\'e de recherches des
 Universit\'es Paris XI et Paris VI associ\'ee au CNRS.},
 Institut de Physique Nucl\'eaire, \\
 91406 Orsay Cedex, France
 \and
         P.-H.~Heenen\thanks{Directeur de Recherches FNRS.} \\
 Service de Physique Nucl\'eaire Th\'eorique,\\
 U.L.B.-C.P.229, B-1050 Brussels, Belgium
 \and
  P.~Bonche \\
SPhT
\thanks{Laboratoire de la DSM} -CE Saclay, 91191
 Gif sur Yvette Cedex, France }
\maketitle

\begin{abstract}
We study the properties of the superdeformed (SD) bands of $^{195}$Pb
and 
$^{193}$Hg by the cranked Hartree-Fock-Bogoliubov method. 
Our calculations reproduce the flat behavior 
of the dynamical moment of inertia 
of two of the SD bands of $^{195}$Pb measured recently. 
We discuss possible configuration 
assignments for the observed 
bands 3 and 4 of $^{195}$Pb. 
We also calculate the two interacting SD bands of $^{193}$Hg. 
Our analysis confirms 
the superiority of a density-dependent pairing force 
over a seniority pairing interaction. 
\end{abstract}

The
dynamical moment of inertia $\cal J$ of most superdeformed (SD) bands 
observed in nuclei of the $A \simeq$ 190 region 
are increasing functions of the angular velocity
$\omega$\cite{ Vo93,Dr91,
Ca90,Fa95,Jo94,Ce94,Fe90,Az91,Az91b,Hu,Hu94,Far95,Wa91,Cl95}. 
The recent discovery by Farris et al.,\cite{Far95} 
that for the two lowest SD 
bands in $^{195}$Pb, $\cal J$ is almost constant versus $\omega$
is therefore particularly interesting.
In the same nucleus two other bands have been observed 
 which display the usual increasing trend.  
It appears natural to attempt an explanation of these various 
behaviors of the SD bands of $^{195}$Pb
in terms of their quasiparticle (qp) structure.
According to most theoretical investigations of the $A=190$ region,
the neutron qp's which are relevant for neutron numbers
above the $N=112$ gap are built on the
[752]5/2, [512]5/2 and [624]9/2 orbitals\cite{Wy95,Gi94,Sa91,Ga94}. 
In this letter, we analyze the properties of the
SD bands of the two odd-$N$ neighbouring nuclei $^{195}$Pb
and $^{193}$Hg \cite{Jo94} which today provide the richest information 
set on the neutron structure in the $A=190$ superdeformed well.
Our work is based on the cranked Hartree-Fock-Bogoliubov (HFB)
approach
which has been shown to reproduce with good accuracy the
SD band properties of even nuclei\cite{Bo96,Te95}. 
As in
ref.~\cite{Ga94}, the mean-field method has been corrected by means of
the Lipkin-Nogami prescription\cite{Li60,No64,Pr73} 
to take into account the finite number of nucleons. 
The nucleon-nucleon effective interaction in the particle-hole channel
is the Skyrme force within the Skm$^*$ parametrization\cite{Ba82}.
In the pairing channel we use a zero-range force with a
surface-peaked density dependence as described in ref.\cite{Te95}. 
In a previous study
of the yrast SD band of $^{194}$Pb, this type of
force was shown to improve significantly the alignment
properties which determine the saturation of $\cal J$
for large values of $\omega$. 

Our method of solution of the HFB equation combines the 
imaginary-time evolution method to determine the basis
which diagonalizes the mean-field hamiltonian and
 a diagonalization of the HFB hamiltonian matrix
to construct the canonical basis. 
Details are available in ref.\cite{Ga94}.
The different bands of the odd nuclei we are concerned with, are
described 
by the self-consistent creation of the appropriate qp \cite{He95} on
the even vacuum.
This requires some care in the numerical treatment, since, except at
zero angular velocity, parity and
signature are the only quantum numbers available
for sorting qp's. 
Therefore, we rely mostly on
continuity properties versus $\omega$ to follow a given SD band. 
The first part of this letter is devoted to a study of $\cal J$ and qp 
routhian (qpr) 
properties of $^{195}$Pb. 
In the second part we analyze how the structure of 
the pairing force affects SD properties by comparing the results 
for $^{193}$Hg with the density dependent interaction 
with those of ref.\cite{He95} in which 
a seniority interaction was used. 

In Fig.~1 we compare the dynamical
moments of inertia $\cal J$ of the four observed
bands\cite{Far95} with those calculated for the seven 
SD bands built on 
the [752]5/2,$\alpha$=$\pm$1/2, [512]5/2,$\alpha$=$\pm$1/2,
[624]9/2,$\alpha$=$\pm$1/2 and [642]3/2, $\alpha$=$-$1/2 orbitals. 
The SD bands built on the intruder 
$\nu$[752]5/2 bands display a small variation of $\cal J$ for
$\hbar\omega$ $\geq$ 0.24 MeV. 
It seems therefore reasonable 
to assign them to the first and second experimental 
bands.
The observed significant signature splitting is also reproduced and
suggest that the band with lowest moment of inertia has a positive
signature.
For bands 3 and 4 the authors of ref.\cite{Far95} have 
argued that they may be built on $\nu$[624]9/2 orbitals. 
On the other hand, the theoretical part of their analysis
indicated that transition energies associated with 
$\nu$[624]9/2 and $\nu$[512]5/2 bands would be almost identical. 
This is supported by our
results.
Indeed we find that the moment of inertia of the $\nu$[624]9/2 bands
is very similar to that of 
the $\nu$[512]5/2 bands for $\hbar\omega$ $\geq$ 0.24 MeV.
Both sets of $\cal J$ agree qualitatively with those observed
for bands 3 and 4. 
Based on the sole information given by the moment of inertia,
it is therefore not possible to decide 
whether bands 3 and 4 
should be labeled $\nu$[624]9/2 or $\nu$[512]5/2.
We only note a small signature splitting for the $\nu$[512]5/2 when 
$\hbar\omega$ $\geq$ 0.3 MeV
A complementary piece of information 
is provided by the excitation 
energy $E_{\rm rel}$ of the bands with respect to each other.
It is generally believed that 
this quantity which is directly available in the
calculation (see Table~1) is correlated with 
the observed relative population of the bands. 
\begin{table}
\centering
\begin{tabular}{cc}
\hline
           band               &  $E_{\rm rel}$ [MeV] \\
\hline
$\nu$[752]5/2, $\alpha$=$+$1/2 &       0.125       \\
$\nu$[752]5/2, $\alpha$=$-$1/2 &       0.000       \\
$\nu$[512]5/2, $\alpha$=$+$1/2 &       0.168       \\
$\nu$[512]5/2, $\alpha$=$-$1/2 &       0.172       \\
$\nu$[624]9/2, $\alpha$=$+$1/2 &       0.297       \\
$\nu$[624]9/2, $\alpha$=$-$1/2 &       0.297       \\
$\nu$[642]3/2, $\alpha$=$-$1/2 &       0.368       \\
\hline
\end{tabular}
\caption{Calculated relative excitation energies $E_{\rm rel}$ of
the six SD bands in $^{195}$Pb at $I$ = 32.5. 
The reference band is $\nu$[752]5/2, $\alpha$=$-$1/2 band. 
$E_{\rm rel}$ of the negative-signature bands were calculated from 
averages of energies of $I$ = 31.5 and 33.5.} 
\end{table}
In Table~1, the reference energy
corresponds to the state of $\nu$[752]5/2, $\alpha$=$-$1/2 band
of the angular momentum $I$ = 32.5, which was calculated by 
averaging two energies of $I$ = 31.5 and 33.5. 
$E_{\rm rel}$ of the other negative-signature bands
were calculated in the same way. 
According to Table~1, the $\nu$[752]5/2 bands are the lowest,
the $\nu$[512]5/2 bands are second lowest and
the $\nu$[624]9/2 bands are the most excited.
Therefore, for the $\nu$[752]5/2 bands, the
agreement of our calculation with experiment
concerns both the magnitude and behavior of moments
of inertia and the excitation energy.
On the other hand, we would be
led to assign a $\nu$[512]5/2 structure to bands 3 and 4.
We note however that the energy differences in Table~1 are
of the order of 0.1MeV.
Such a precision could well be below
the limit of physical credibility of a calculation such as ours
when it comes to the relative position of orbitals.
We shall return to this point
when discussing the crossing phenomenon in $^{193}$Hg.
 
The charge quadrupole moments $Q_{\rm c}$
of the $\nu$[512]5/2 and $\nu$[624]9/2 bands
are shown in Fig.~2.
They differ by approximately 0.3b 
which is 
probably too small to be measured.  
As expected for bands with large $m$ values 
no signature splitting is found. 
The quadrupole moments of the intruder bands are
more different: 19.96 eb at $I$ = 32.5 ($\hbar\omega$ = 0.314MeV )
for  $\alpha=+1/2$ and  19.47 eb at $I$ = 31.5
( $\hbar\omega$ = 0.290MeV )
for its signature partner.
Differences between the magnetic moments of the
$\nu$[512]5/2 and $\nu$[624]9/2 bands
would lead to different crosstalks
between the bands.
However their values are rather similar ( 
$\simeq$ 12.4 $e\hbar/2M_{\rm p}c$ at $I$ $\simeq$ 31, $M_{\rm p}$
being the 
proton mass ) 
for the four bands.
They therefore also do not provide a convenient signature to 
establish the nature of the 3rd and 4th SD bands. 

Let us now consider the evolution versus $\omega$
of the qpr's. Because the mean-fields are self-consistently
modified by the creation of a qp, they are not the same
for the six SD bands and differ also from those
calculated for $^{194}$Pb.
In Fig.\ 3 and 4 we show the neutron qpr's for the $\nu$[752]5/2,
$\alpha$ = $-$1/2 
( after a crossing with [512]5/2,$\alpha$=$-$1/2 )
and $\nu$[624]9/2, $\alpha$ = +1/2 bands
respectively.
In each figure, the thick curve indicates which qp is 
occupied. 
One sees that the difference between the occupied and empty qpr's 
of the signature partners is large for all values of the rotational
frequency. 
This is caused by the modification of the time-odd components
of the mean-field that is generated by the occupation of
only one of the signature partner ;  
in particular one notes that the difference does not vanish 
at $\omega$ = 0. 
Figs.\ 3 and 4 show that the flat $\cal J$ behavior 
for the $\nu$[752]5/2 
bands is strongly correlated with the curvature of the associated
routhians which is markedly different from
those of other qpr's. 
Moreover, the fact that the average curvature of the
$\nu$[752]5/2, $\alpha$ = $+$1/2 routhian is 
the smallest, is consistent
with the low moment of inertia of the corresponding band.

The accident in the theoretical $\cal J$ for
the [752]5/2 and [512]5/2 bands
for $0.1$MeV$\leq\hbar\omega\leq$0.2MeV
is generated by a band crossing.
Such an accident can always 
happen when two crossing bands have the same quantum numbers.
In case of a band crossing, our convention is to denote bands with
the qp configuration which characterizes them for
large values of $\hbar\omega$.
For values of $\hbar\omega$ near 0.15 MeV,
both for the $\nu$[752]5/2 and $\nu$[512]5/2 bands,
we have not been able to 
obtain solutions satisfying the angular-momentum constraint
accurately.
Indeed, we have met numerical instabilities
caused by the near degeneracy both in energy
and angular momentum.
A correct physical solution of this problem
 requires a self-consistent configuration mixing calculation
which is beyond the scope of this study.
As a band crossing has not been observed in $^{195}$Pb,
we infer that the calculation shown in Fig.~3 
overestimates the energy difference between the $\nu$[752]5/2, 
$\alpha$=$-$1/2 
and $\nu$[512]5/2, $\alpha$=$-$1/2 energies at $\omega=0$
by at least 0.1MeV.
One way  to remedy this deficiency,
would be to correct the mean-field in such a way that 
the energy of the $\nu$[512]5/2, $\alpha$=$-$1/2
is pushed up, leading to a crossing below the lowest observed
$\hbar\omega$.
According to the above discussion, 
the associated  excitation 
energy of the $\nu$[512]5/2 SD bands
would increase and possibly modify our assignation for band 3 and 4, 
leading us to agree with the conclusions of ref.\cite{Far95}.
 
We should also mention the relative position of $\nu$[624]9/2,
$\alpha$=+1/2
in Fig.~4. 
The qpr is not the lowest in routhians having the positive parity and 
positive signature. 
It is anticipated, however,  that a particle-type qpr ($\nu$[624]9/2) 
becomes lower than a hole-type one ($\nu$[642]3/2) in $^{195}$Pb, when 
their energies are comparable for the yrast SD band of $^{194}$Pb. 
Given the present uncertainty of mean-field calculation concerning
the detailed relative location of the qpr's,
we cannot disregard $\nu$[624]9/2 as one of 
candidates of the configuration of bands 3 and 4.

 On Fig.\ 2 and in Table 1 we have also reported result for 
the negative signature band built on 
the $\nu$[642]3/2,$\alpha$=$-$1/2 quasiparticle. 
Although the routhian of this state is lower than that of 
the $\nu$[624]9/2 in the quasiparticle spectrum of $^{194}$Pb, 
Table 1 shows that the [642]3/2 SD band is more excited. 
This is a consequence of self-consistency effects. 

Let us now see how these considerations can be extended to
the analysis of the nucleus $^{193}$Hg. 
So far 6 SD bands including two identical bands have 
been observed \cite{Jo94}. 
Within the HFB method we have already performed a study 
using the same Skyrme force parametrization for the
mean-field\cite{He95}.
On the other hand, in ref.\ \cite{He95} pairing correlations
have been described with a seniority interaction. 
The results 
of analysis limited to the
two interacting bands $\nu$[752]5/2, $\alpha$ = 
$-$1/2 band (band 4) and $\nu$[512]5/2, $\alpha$ = $-$1/2 
band (band 1) with a zero-range density dependent
pairing force are shown in Fig.\ 5
together with the experimental data. 
For each band a separate HFB calculation has been performed.
Our calculation shows that this dual self-consistent
HFB analysis is able to reproduce the observed band interaction.
This is a significant improvement over the calculation
of ref.~\cite{He95} in which no interaction was found, 
and it provides an additional indication of the superiority of
a surface-type zero-range pairing force over a seniority interaction. 
On the other hand, the angular velocity at which 
bands interact is found at $\hbar\omega\simeq0.15$MeV
instead of the observed value $\hbar\omega\simeq0.25$MeV. 
This difference may reflect an inaccuracy of the relative location of 
the relevant qpr's in $^{193}$Hg which would then be consistent with
our discussion on the position of orbitals 
[752]5/2 and [512]5/2 in $^{195}$Pb. 

In summary, we have analyzed the properties of the SD bands of two 
odd-$N$ nuclei $^{195}$Pb and $^{193}$Hg by making use of the cranked 
HFB method. 
Our self-consistent calculation has confirmed the general belief that 
the flat behavior of $\cal J$ in bands 1 and 2 of $^{195}$Pb is
related 
to the curvature of the neutron intruder qpr. 
For bands 3 and 4 we have found that configurations based on 
the [624]9/2 and [512]5/2 are in competition. 
Both the moment of inertia, the quadrupole moments 
and the magnetic moments of these four SD bands are very similar.
In particular, within the HFB method these quantities do not provide
an effective mean of deciding the nature of the observed bands 3 and
4. 
We have also calculated with qualitative success
the two interacting SD bands of $^{193}$Hg. 
The results of this analysis provide additional support for an 
effective pairing force acting predominantly at the nuclear surface. 
The quantitative inaccuracy on the position of the crossing
frequencies 
($\Delta\hbar\omega\approx$ 0.1 MeV) could be an indication that 
qpr associated with the [512]5/2 is too low by 0.1 MeV 
relative to the rest of the spectrum. 
It is an interesting question 
whether it is possible to determine an effective interaction
with the same global qualities of the SkM$^{\ast}$ force
which could also achieve a better precision
as regards the single-particle energies.

{\bf Acknowledgements}

We thank J.~Becker and L.~Farris for discussions on 
the experimental results. 
This work has been partly supported by the ARC convention 
93/98-166 of the Belgian SSTC.

\newpage
\noindent
Figure Captions

\noindent
\begin{itemize}
\item[Fig.~1] a) Experimental dynamical moments of inertia of the four 
SD bands \cite{Far95} of $^{195}$Pb. 

b) HFB dynamical moments of inertia of the lowest SD bands of
$^{195}$Pb. 
( the negative signature bands are denoted with black symbols ) 

\item[Fig.~2] Calculated charge quadrupole moments $Q_{\rm c}$ of 
the $\nu$[512]5/2, $\nu$[624]9/2 bands and the $\nu$[642]3/2,$\alpha$
= $-$1/2
band of $^{195}$Pb. 
Correspondence between the symbols and the SD bands is the same as in
Fig.~1b. 

\item[Fig.~3] Neutron qpr of $\nu$[752]5/2, $\alpha$=$-$1/2 SD band. 
The thick dotted curve indicates the occupied qpr. 
Full (dashed) curves correspond to positive parity and positive
(negative) 
signature, and dot-dashed (dotted) curves to negative parity and 
positive (negative) signature. 

\item[Fig.~4] Neutron qpr of $\nu$[624]9/2, $\alpha$=+1/2 SD band. 
The thick full curve corresponds to the occupied qpr. 
The other drawing conventions are the same as those used in Fig.\ 3. 

\item[Fig.~5] Experimental dynamical moment of inertia $\cal J$ of 
$^{193}$Hg for  bands 1 (solid triangle) and 
 4 (solid circle).
Our results are indicated by open triangles for the $\nu$[512]5/2,
$\alpha$=$-$1/2
configuration and by open circles for the $\nu$[752]5/2, $\alpha$=$-$1/2
one.

\psfig{figure=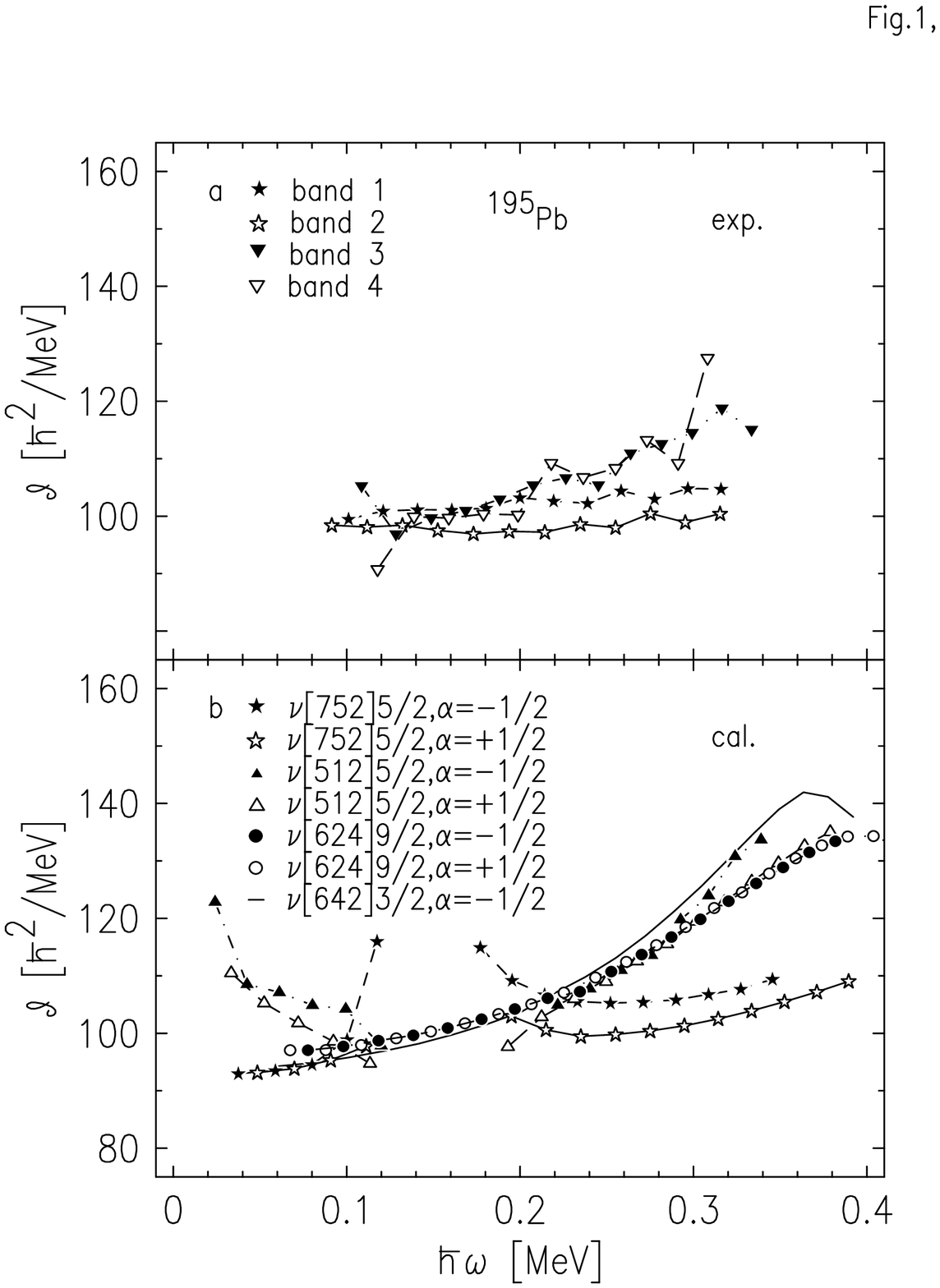}
\psfig{figure=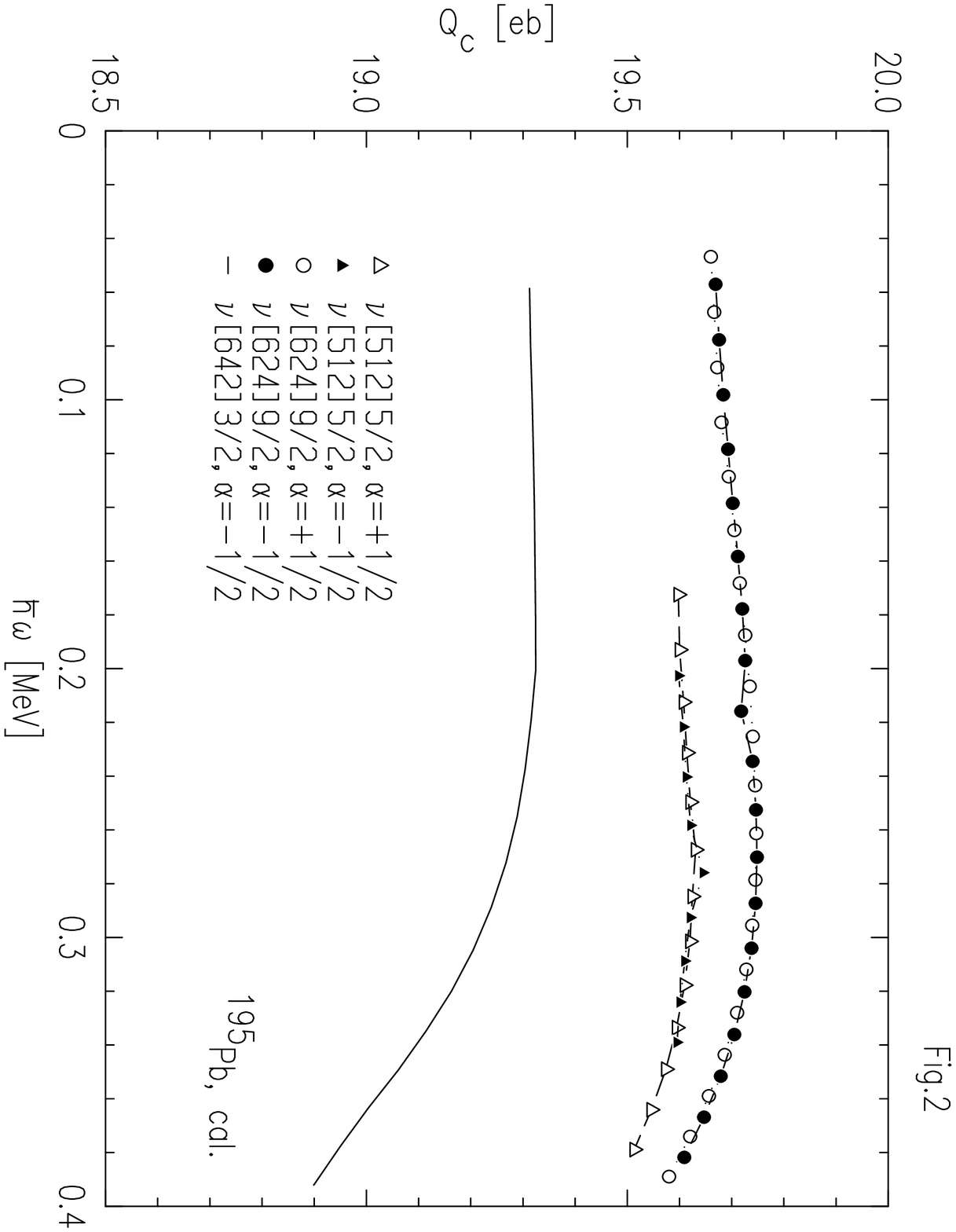}
\psfig{figure=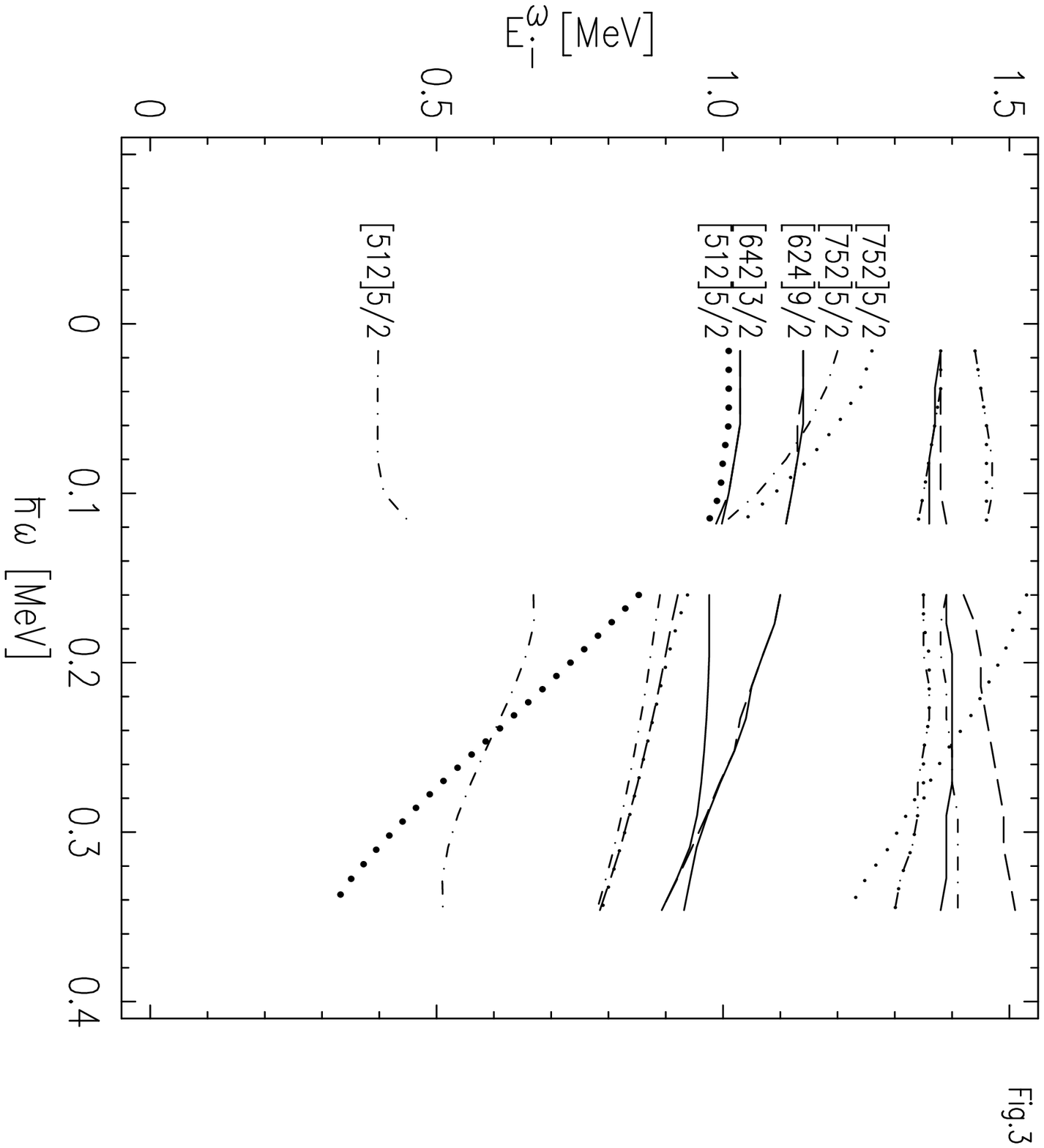}
\psfig{figure=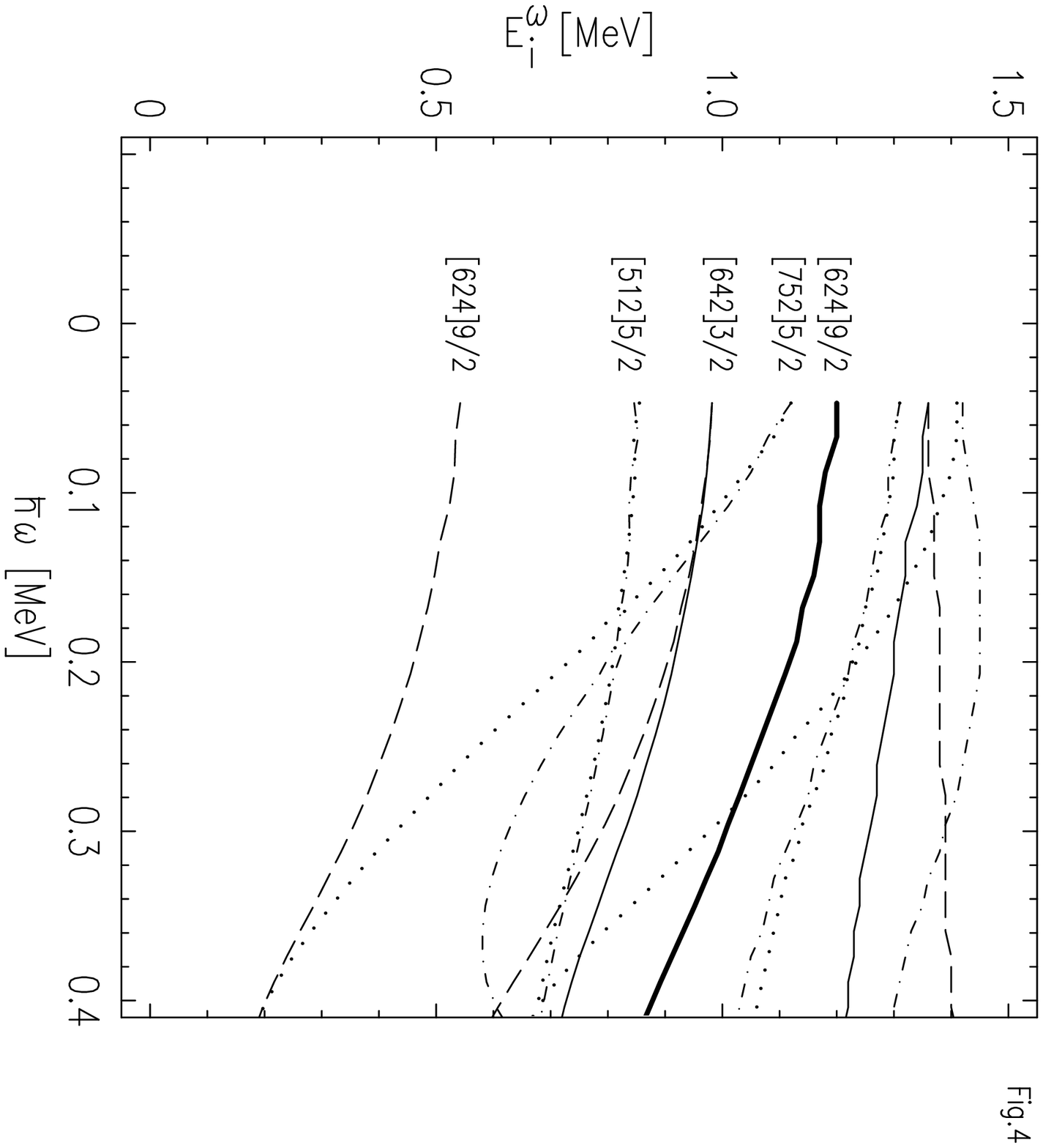}
\psfig{figure=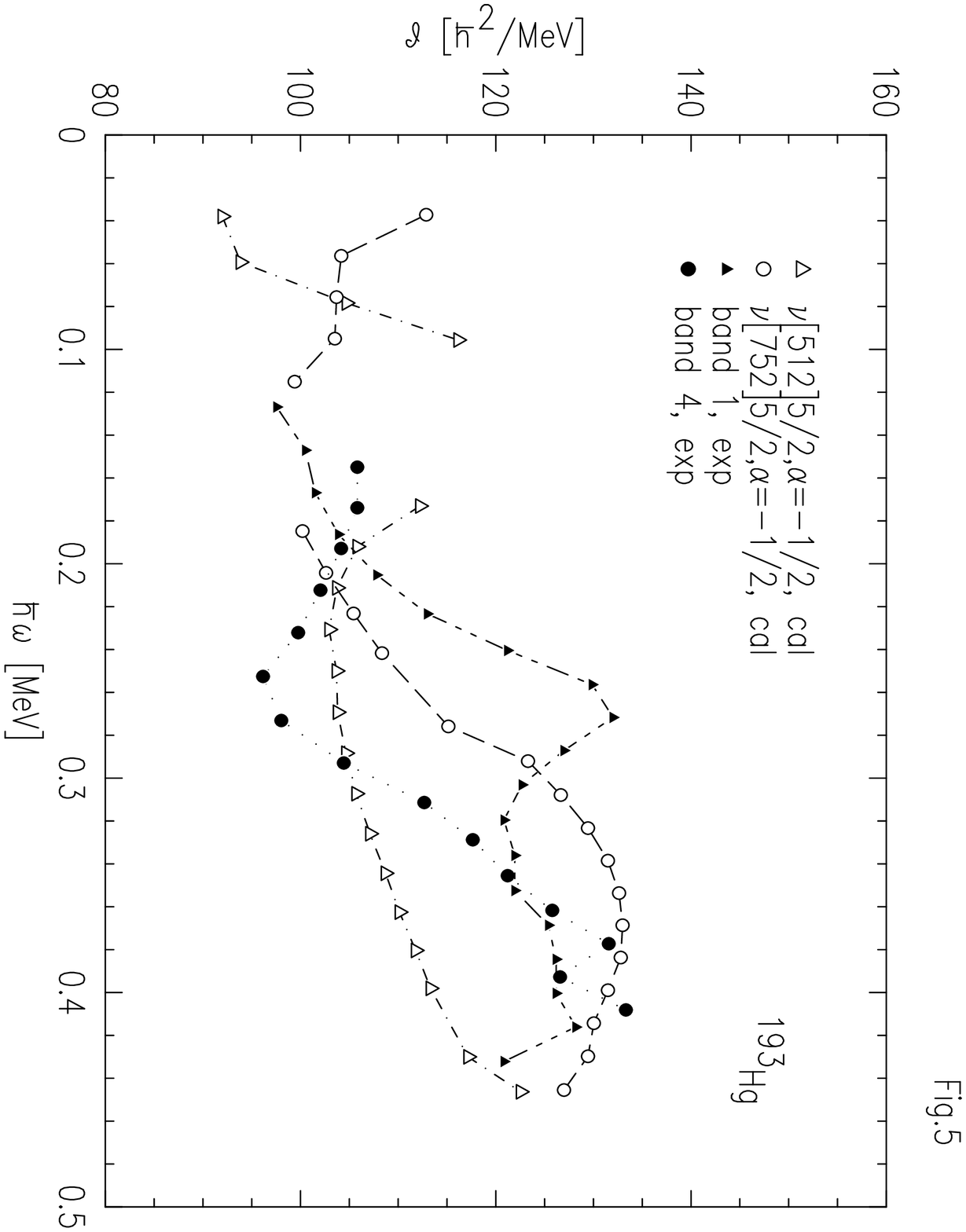}
\end{itemize}

\begin{thebibliography}{99}
%
\bibitem{Vo93} D.\ T.\ Vo et al., Phys.\ Rev.\ Lett.\ {\bf 71} (1993)
340
%
\bibitem{Dr91} M.\ W.\ Drigert et al., Nucl.\ Phys.\ {\bf A530} (1991)
452
%
\bibitem{Ca90} M.\ P.\ Carpenter et al., Phys.\ Lett.\ B {\bf 240}
(1990) 44
%
\bibitem{Fa95} P.\ Fallon et al., Phys.\ Rev.\ C {\bf 51} (1995) R1609
%
\bibitem{Jo94} M.\ J.\ Joyce et al., Phys.\ Lett.\ B {\bf 340} (1994)
150
%
\bibitem{Ce94} B.\ Cederwall et al., Phys.\ Rev.\ Lett.\ {\bf 72}
(1994) 3150
%
\bibitem{Fe90} P.\ B.\ Fernandez et al., Nucl.\ Phys.\ {\bf A517}
(1990) 386
%
\bibitem{Az91} F.\ Azaiez et al., Phys.\ Rev.\ Lett.\ {\bf 66} (1991)
1030
%
\bibitem{Az91b} F.\ Azaiez et al., Z.\ Phys.\ A {\bf 338} (1991) 471
%
\bibitem{Hu} J.\ R.\ Hughes et al., Phys.\ Rev.\ C
%
\bibitem{Hu94} J.\ R.\ Hughes et al., Phys.\ Rev.\ C {\bf 50} (1994)
R1265
%
\bibitem{Far95} L.\ P.\ Farris et al., Phys.\ Rev.\ C {\bf 51} (1995)
R2288
%
\bibitem{Wa91} T.\ F.\ Wang et al., Phys.\ Rev.\ C {\bf 43} (1991)
R2465
%
\bibitem{Cl95} R.\ M.\ Clark et al., Phys.\ Rev.\ C {\bf 51} (1995)
R1052
%
\bibitem{Wy95} R.\ Wyss and W.\ Satu{\l}a, Phys.\ Lett.\ B {\bf 351}
(1995) 393
%
\bibitem{Gi94} M.\ Girod, J.\ P.\ Delaroche, J.\ F.\ Berger and
J.\ Libert, 
	       Phys.\ Lett.\ B {\bf 325} (1994) 1
%
\bibitem{Sa91} W.\ Satu{\l}a, S.\ \'Cwiok, W.\ Nazarewicz, R.\ Wyss
and 
	       A.\ Johnson, Nucl.\ Phys.\ {\bf A529} (1991) 289
%
\bibitem{Ga94} B.\ Gall, P.\ Bonche, J.\ Dobaczewski, H.\ Flocard,
P.-H. Heenen,
	       Z.\ Phys.\ A {\bf 348} (1994) 183
%
\bibitem{Bo96} P.\ Bonche, H.\ Flocard and P.-H.\ Heenen, 
	       Nucl.\ Phys.\ {\bf A598} (1996) 169
%
\bibitem{Te95} J.\ Terasaki, P.-H.\ Heenen, P.\ Bonche,
J.\ Dobaczewski and 
	       H.\ Flocard, Nucl.\ Phys.\ {\bf A593} (1995) 1
%
\bibitem{Li60} H.\ J.\ Lipkin, Ann.\ Phys.\ (N.\ Y.\ ) {\bf 9}
(1960) 272
%
\bibitem{No64} Y.\ Nogami, Phys.\ Rev.\ {\bf 134} (1964) B313
%
\bibitem{Pr73} H.\ C.\ Pradhan, Y.\ Nogami, J.\ Law, Nucl.\ Phys.\ 
{\bf 201} (1973)
               357
%
\bibitem{Ba82} J.\ Bartel, P.\ Quentin, M.\ Brack, C.\ Guet and
H.-B.\ H{\aa}kansson,
               Nucl.\ Phys.\ {\bf A386} (1982) 79
%
\bibitem{He95} P.-H.\ Heenen, P.\ Bonche and H.\ Flocard, 
	       Nucl.\ Phys.\ {\bf A588} (1995) 490
%
\end{thebibliography}
\end{document}